\documentclass[%
 aip,
 sd,%
 amsmath,amssymb,
 reprint,%
]{revtex4-1}

\usepackage{graphicx}
\usepackage{dcolumn}
\usepackage{bm}
\usepackage{amsmath}

\begin{document}

\title{Wave Physics of  the Graphene Lattice Emulated in a Ripple Tank}

\affiliation{Solid State Physics Laboratory, ETH Z\"{u}rich, 8093 Z\"{u}rich, Switzerland}
\author{J. R\"{o}ssler}

\author{ C. R\"{o}ssler}

\author{ P. M\"{a}rki}
\author{ K. Ensslin}
\author{ T. Ihn}

\date{\today}

\begin{abstract}

Using the example of graphene, we have extended the classic ripple tank  experiment to illustrate the behavior of waves in periodic lattices.
A loudspeaker driving  air through a periodically perforated plexiglass plate onto the tank's water surface creates wave patterns that are in agreement with numerical simulations and are explained in terms of solid state theory. 
From an educational point of view, the experiment provides an illustrative example of the concepts of reciprocal space and symmetry. 
\end{abstract}

\maketitle

\begin{figure}[h]
\includegraphics{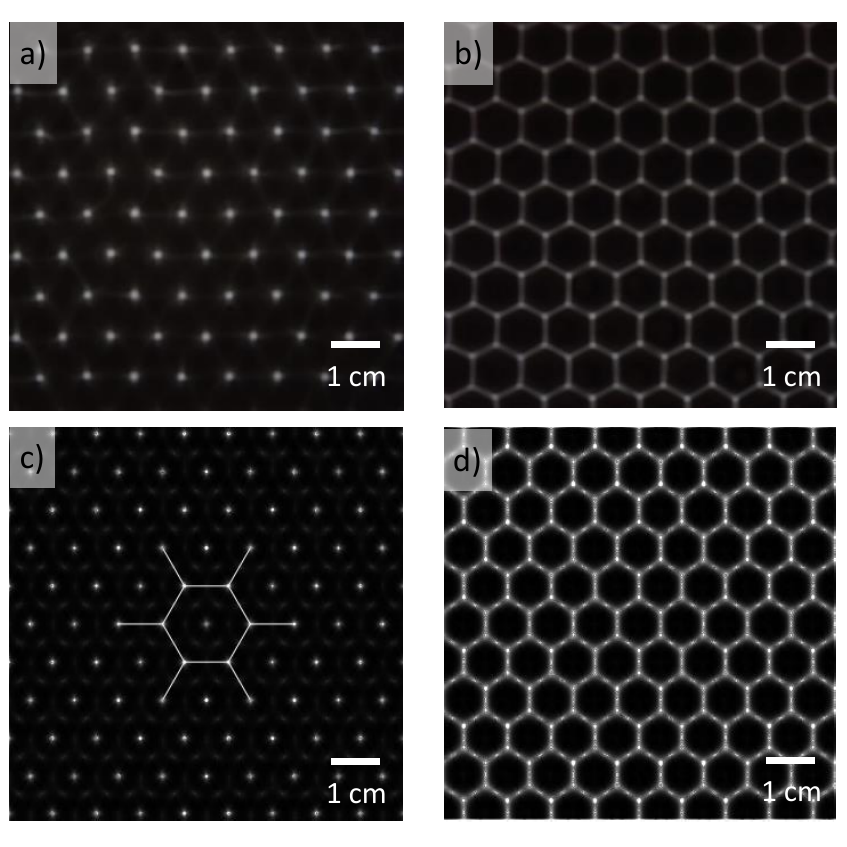}
\caption{\label{fig:hexagon}  $\bf{(a)}$: Photograph of the observation screen picturing the ripple tank's surface while driving air through a plexiglass plate with  660 hexagonally arranged holes.  The next  neighbor distance is a$_0$ = 1 cm. The excitation frequency that generates the sharpest patterns corresponds to a wavelength in water of  $\lambda =\sqrt{3}/2\,\rm{cm}$. 
$\bf{(b)}$: like in $\bf{(a)}$, but with an additional phase difference between stroboscope and excitation of  $\pi$. The centers of the dark hexagons correspond to the bright spots in $\bf{(a)}$. Over time, the wave oscillates back and forth between these patterns.
$\bf{(c/d)}$: Simulation of the 2D wave pattern obtained by solving the wave equation for  660 time-harmonic sources in a water-like environment, see Eq. \ref{hankel}. Parameters like in $\bf{(a)}$  and $\bf{(b)}$, respectively.  The white lines in $\bf{(c)}$ indicate the honeycomb pattern of the source plate.}
\end{figure}

At public outreach events as well as in schools or undergraduate lectures, the ripple tank  is widely used to illustrate wave phenomena.\cite{kuw85,log14}
However,  a standard ripple tank's one or two dappers cannot create patterns that illustrate the behavior of waves in periodic lattices. This challenged us to modify the setup in a way that provides multiple periodic sources. Employing a loudspeaker and a plexiglass plate with hexagonally arranged holes we were able to realize  wave patterns in the ripple tank, that are reminiscent of the microscopic images of artifical graphene.\cite{man12} Two characteristic interference pattens are depicted in  Figs. \ref{fig:hexagon}(a) and (b). Readers specialized in solid state physics will recognize the same symmetry as graphene electronic wave functions in the center of the first Brillouin zone. We find  that the strongest interference patterns are not  obtained when the excitation wavelength is equal to the nearest neighbor separation $a_0 $. Instead, the symmetry of the lattice favors wavelengths that are 13.4 $\%$ shorter than $a_0 $. The discrepancy hints at the non-trivial symmetry properties of the honeycomb lattice and is explained by means of solid state theory, testifying to the applicability of our ripple tank for educational purposes.

\begin{figure}[hhh]
\includegraphics{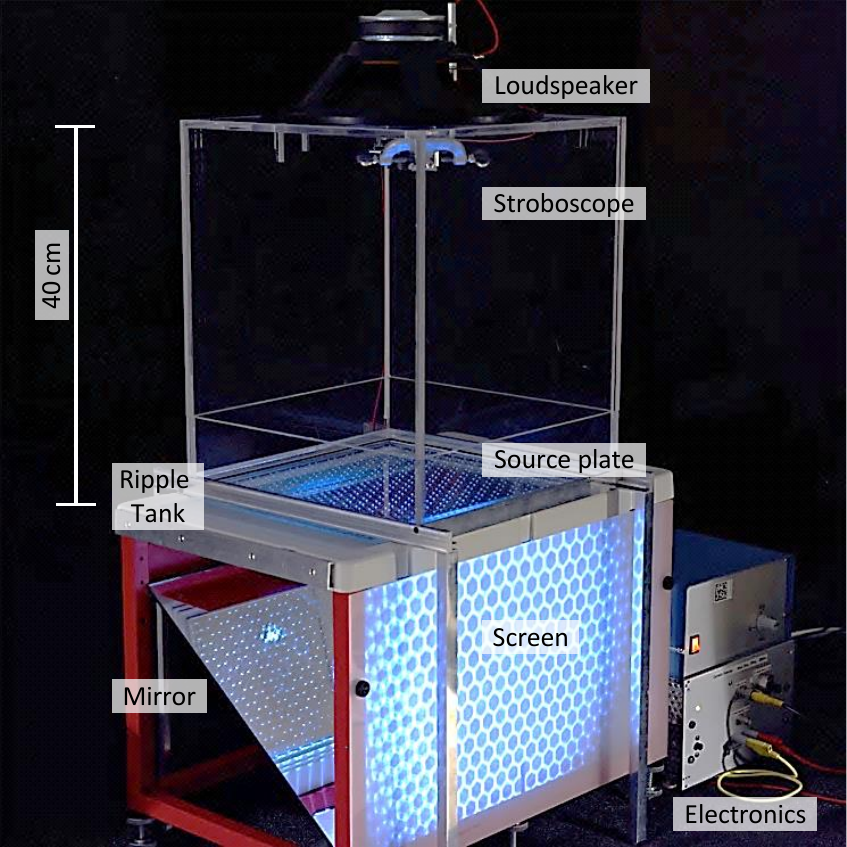}
\caption{\label{fig:Setup} Setup of the ripple tank showing a 2D honeycomb  wave pattern. Sound waves travel from the loudspeaker (top) through the hexagonally patterned source plate onto the ripple tank's surface. The image is created by a stroboscopic LED (top) that projects the wave pattern via the mirror onto the screen.}
\end{figure}

The paper is organized as follows:
section \ref{Setup} describes the technical details of the setup, in section \ref{singleSource} and \ref{twoSources} we measure its physical properties, and in section \ref{theoryHex} the favorable excitation wavelengths are extracted from the discrete Fourier transform of a finite honeycomb lattice. Section \ref{correspondence} targets the correspondence between water waves in the ripple tank and electronic wave functions in graphene. We close with summary and outlook. 

\section{\label{Setup}Setup}
Figure \ref{fig:Setup} gives an overview of the modified ripple tank setup.
A loudspeaker  is used as a  precise and inexpensive source of excitation.\cite{gre77}
It is mounted face-down inside a plexiglass hood. An aluminium rack holds both the hood and an interchangeable  source plate  295 $\times$ 295 $\times$ 4.2 mm$^3$ in size. The source plate seen in  Fig.~\ref{fig:Setup}  is patterned with a honeycomb lattice of holes.  Directly underneath the source plate we place a commercially available ripple tank (Leybold,  No. 401 501, with  see-through water trough,  tilted mirror and observation screen). The distance between source plate and water surface is only a few mm.  A white LED  mounted in front of the center of the loudspeaker illuminates the ripple tank surface and can be used as stroboscope to observe a wave pattern at fixed phase.  To that end, the LED is typically powered  for 1 $\%$ of time of each wave cycle, thus allowing the observation of the wave during a 4$^{\circ}$ phase interval.

Both the stroboscope's  and the loudspeaker's  frequencies, relative phase-shift, wave form and amplitude are controlled with a LabView program  on a standard PC via an USB Audio Adapter (LogiLink, UA0078).  Between PC and loudspeaker or stroboscope, amplifiers are used to scale the signals appropriately. The loudspeaker requires between 0.3 and 3V AC input voltage, the LED 4V DC. The loudspeaker receives a sinusoidal input signal, the stroboscope's waveform is set to rectangular. 

In combination with the loudspeaker, the hole pattern of the source plate thus creates a controlled spatial pattern of point sources of sinusoidal-in-time airblasts directed downward onto the surface of the water.\cite{bal30} Since ripple crests/valleys act on the illuminating light as converging/diverging lenses,\cite{gre12} the resulting wave pattern on the water surface becomes visible as bright and dark areas projected onto the observation screen.  

\section{\label{singleSource}Single source}

\begin{figure}[h]	
\includegraphics{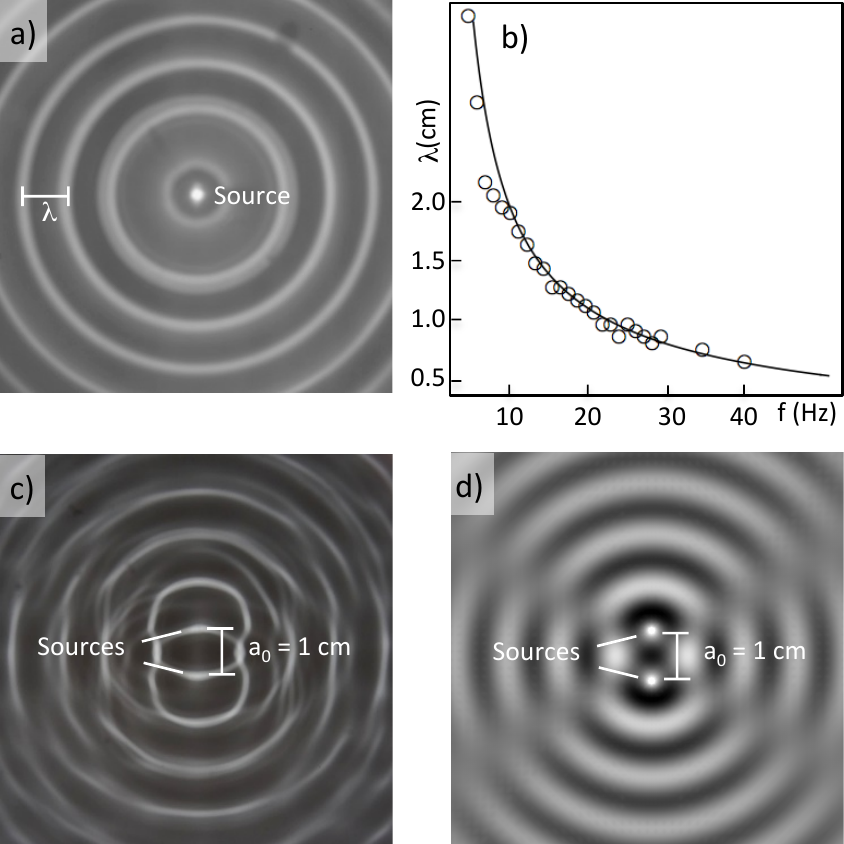}
\caption{\label{fig:Dispersion} $\bf{(a)}$: Photograph of the  observation screen for a single point source plate at $\lambda$ = 1 cm. $\bf{(b)}$:  Measured (circles) and calculated (line) dispersion relation in the ripple tank. $\bf{(c)}$: Photograph of the observation screen for a double point source plate at  $\lambda$ = 1 cm. $\bf{(d)}$: Simulation of the 2D wave pattern created by two time-harmonic sources in a water-like environment.}
\end{figure}

We  performed a series of single source experiments to  determine the dispersion relation of  water waves in our setup.  
The water height in the tank was kept constant  at $h = 1.8$ cm. The wavelength lies  between 0.5 cm and 2.5 cm. 
Since wavelength and water height are of the same order of magnitude, the dispersion relation for ripples is to a good approximation \cite{cla59} given by
\begin{equation}
\omega^2 =gk+ \frac{S k^3}{\rho}
\label{dispersion} 
\end{equation}
where $\omega$ is the angular frequency of the ripple, $k = 2 \pi/\lambda$ the wavevector, $\rho$ the density of the water ($1\,\rm{g/cm^3}$), $S$ the surface tension ($73 \,\rm{g/s^2}$),  and $g$ the absolute value of the gravitational acceleration ($981 \,\rm{cm/ s^{2}}$). 

For single source experiments, we fabricated a source plate with a single hole 2 mm in diameter.   The frequency was swept from 5 to 40 Hz. With a digital camera, we took pictures of the observation screen. The radial difference between two adjacent bright circles in  Fig.~\ref{fig:Dispersion}(a)  corresponds to the wavelength.   Fig.~\ref{fig:Dispersion}(b)  shows our measured wavelengths  in comparison to Eq. \ref {dispersion}. 
For frequencies above 10 Hz, we find a good agreement, The deviation at low excitation frequencies could be caused by operating the loudspeaker outside the specified range which might introduce nonlinearities in the excitation.
Fitting the data using least mean squares, we extract a measurement uncertainty of  $\sigma = $1\, Hz for each data point, assuming that all uncertainties are statistical in nature. 
Measurement uncertainties comprise calibration errors of the gauge scale on the observation screen, reading errors of the wavelength and calibration errors of the frequency generator.
Thus, for example at 30 Hz,  the wavelength of the water waves generated by our setup is known with $0.02$\, cm precision.

\section{\label{twoSources}Two sources}

Using a  plate with two holes, we can emulate a double slit experiment. 
We choose this well-known geometry to  compare measurement with simulation as well as to study the effect the finite hole diameter has on the pattern.

The angles of constructive and destructive interference\cite{poo02} can be read from  Fig.~\ref{fig:Dispersion}(c). 
 Starting with the solution to the wave equation of  a 2D time-harmonic source \cite{wat14} for a radially propagating wave

\begin{equation}
\psi (r, k, \omega, t) = \frac{-i}{2 \pi\sqrt{2}} ( H_0^{2}[k r ]  e^{i \omega t} + H_0^{2}[-k r ]  e^{-i \omega t} )
\label{hankel} 
\end{equation}
with $H_0^{2}$ the zero order Hankel function of the second kind, $k$ the absolute value of the wave vector,  $r$ the radial distance from the source, $\omega$ the radial frequency,  and $t$ the time,
we performed a Mathematica-based simulation of the 2D wave pattern. The result of these calculations  is shown in  Fig.~\ref{fig:Dispersion}(d). 

To estimate the effect of the finite hole size, we compare a simulated wave function $\psi_1(x,y)$ with two point-like sources  to a wave function $\psi_2(x,y)$, where each source is replaced by five sources at the four endpoints and the center of a crosshair whose diameter corresponds to the diameter of the holes.
As a figure of merit, we introduce the correlation between $\psi_1(x,y)$ and $\psi_2(x,y)$ given by: 
\begin{equation}
\zeta = \frac{\left< \psi_1(x,y) \,\psi_2(x,y)\right> }{\sqrt{\left< \left|\psi_1(x,y)\right|^2\right>\left< \left|\psi_2(x,y)\right|^2\right>} }
\label{correlation} 
\end{equation}
Using  a distance between the sources of $a_0 =  1 \,\rm{cm}$ and a hole diameter of 2 mm we find  $\zeta$ to be 99.7  $\%$. 
Since these parameters are also convenient to realize experimentally, we use them with confidence in all subsequent source plates.

\section{\label{theoryHex}Honeycomb lattice}

The symmetry properties of the honeycomb lattice  can be inferred by performing the  Fourier transform of the wave pattern. We exploit that this transform can be expressed as the product of the Fourier transform of a single time harmonic point source and that of the lattice. We perform a discrete Fourier transform of a finite lattice and compare the results to key parameters known from the Fourier transform of an infinite lattice,  which are deduced from the pseudopotential method for a crystal with a two-atom basis.\cite{ihn10}

 \begin{figure}[h]
\includegraphics{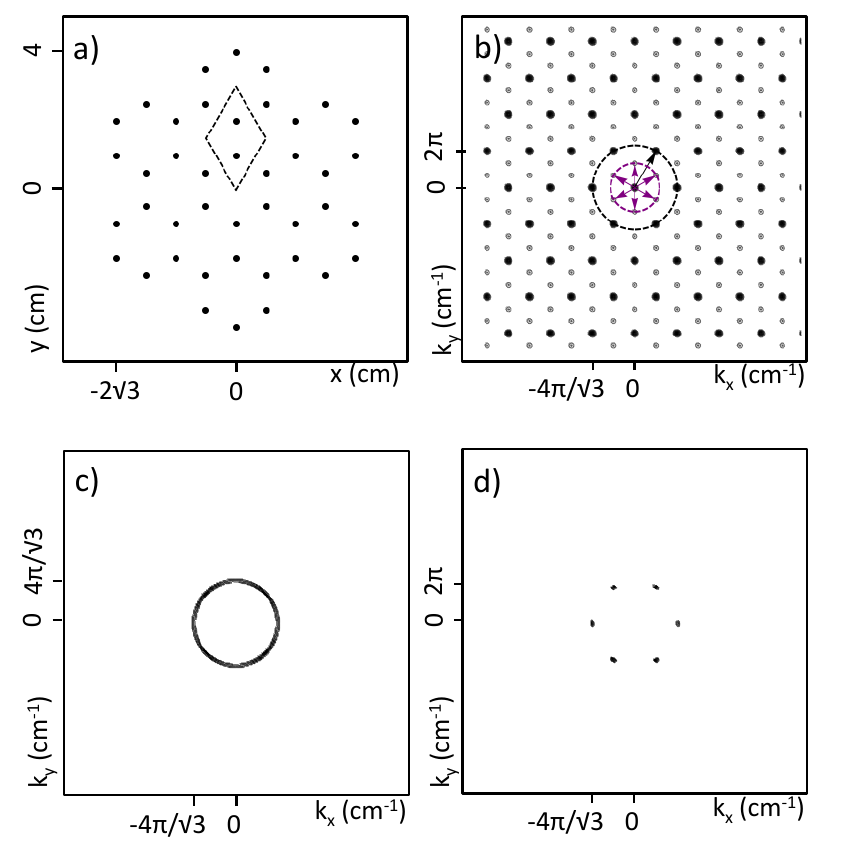}
\caption{\label{fig:Fourier} $\bf{(a)}$: Honeycomb pattern with 42 holes: nearest-neighbor distance $a_0$ is 1 cm.  Dashed black lines indicate the unit cell of the lattice. $\bf{(b)}$:  2D discrete Fourier transform of the pattern depicted in $\bf{a}$. Dark spots correspond to high amplitude. Purple arrows indicate the six lowest reciprocal lattice vectors, the black arrow indicates the next highest reciprocal lattice vector. $\bf{(c)}$: 2D discrete Fourier transform of a single time harmonic source with wavelength $\lambda = \sqrt{3}/2\,\rm{cm}$.  $\bf{(d)}$: 2D discrete Fourier transform of 42 time harmonic point sources with wavelengths of  $\lambda = \sqrt{3}/2\,\rm{cm}$.}
 \end{figure}

We created  a 42 point lattice with $a_0$ = 1 cm for the discrete transform.
Figure \ref{fig:Fourier}(a) pictures this lattice in real space. 
Figure \ref{fig:Fourier}(b) shows the squared modulus of its discrete Fourier transform. As expected from the pseudopotential method we find a pattern with two classes of points with an intensity ratio of 1 to 4. 
Purple arrows indicate the 6 shortest reciprocal lattice vectors, $\vec{b}_1$, $\vec{b}_2$, $\vec{-b_1}$, $\vec{-b_2}$, $\vec{b}_1-\vec{b}_2$ and $\vec{-b_1 }+\vec{ b_2}$, with 

\begin{equation}
\vec{b}_1  =  \frac{2\pi}{a_0\,\sqrt{3}} \begin{pmatrix} 1 \\  1/\sqrt{3} \end{pmatrix}   \, \,\,,\,\, \vec{b}_2  = \frac{2\pi}{a_0\,\sqrt{3}} \begin{pmatrix} 0 \\ 2/\sqrt{3} \end{pmatrix} 
\end{equation}

\noindent and  correspond to the weak intensity points in the pattern. The black arrow indicates the next longer reciprocal lattice vector $\vec{b}_1 + \vec{b}_2$ .  The radii of the dashed circles in pink and black correspond to the $k$-vector lengths favoured by the pattern, while all other lengths are suppressed. 
The finite lattice size employed in the calculation manifests itself in the width of the points.

The discrete 2D Fourier transform of a single time harmonic point source is shown in   Fig.~\ref{fig:Fourier}(c) for the specific wavelength of $\sqrt{3}/2\,\rm{cm}$. We see a circle of radius $4 \pi/\sqrt{3}\,\rm{cm^{-1}}$. 

The final result, the 2D Fourier transform of the  42 time harmonic point sources is shown in  Fig.~\ref{fig:Fourier}(d), which pictures the product of the Fourier transforms shown in  Fig.~\ref{fig:Fourier}(b)  and  Fig.~\ref{fig:Fourier}(c).

Consequently, the discrete 2D Fourier transform of  Fig.~\ref{fig:hexagon}(a) which was taken at  an excitation wavelength of $\lambda = \sqrt{3}/2\,\rm{cm}$ yields six points on a circle  in k-space, analogous to  Fig.~\ref{fig:Fourier}(d).

This analysis illustrates that the resulting wave pattern in real space is, at the given wave length, essentially a superposition of six plane wave states. In general, the excitation wavelength, which determines the radius of the circle in   Fig.~\ref{fig:Fourier}(c), therefore serves as a tool to select the finite set of plane wave states that superimpose in real space. Most importantly,  it is not the correspondence between the nearest neighbor separation and the wave length in real space that gives rise to strong interference patterns, as one could naively think. Instead, the wavelength of strongest interference $\lambda = \sqrt{3}/2\,a_0$ corresponds to half the side length of the rhombus as indicated by the black lines in Fig.~\ref{fig:Fourier}(a). Thus, the lattice ripple tank experiment leads to the conclusion that the rhombus and not the honeycomb is the unit cell of the graphene lattice.\cite {wal47} The rhombus is the smaller and thus more fundamental unit that any honeycomb lattice can be constructed from.  

\section{\label{correspondence}Correspondence to graphene}

In the present experimental setting, all  sources are in phase, since they are excited by a common air pulse. This means that as far as the correspondence to graphene is concerned, we are limited to mimicking wave functions consisting of counterpropagating waves at the $\Gamma$- point.

\begin{figure}[h]
\includegraphics{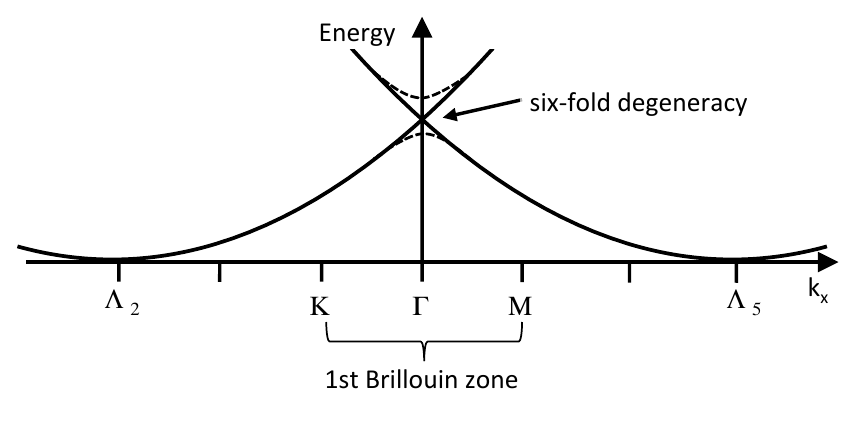}
\caption{\label{fig:dispersion} Cut through the 2D bandstructure of graphene along the $\vec{k}_x$ direction (schematic drawing). On the x-axis, the symmetry points and the first Brillouin zone  are indicated.  The dashed lines indicate the lifting of the degeneracy due to the atomic potential.}
 \end{figure}
The strongest resonance  in the ripple tank is excited with
$\lambda$ = $\sqrt{3}/2 \, a_0$. As discussed in section \ref{theoryHex}, it  corresponds to  $k$ -vectors  of length $k =  \frac{4 \pi}{\sqrt{3} \, a_0}$. There is a superposition of six basis vectors of the reciprocal lattice with this k-vector length,  see  Fig. ~\ref{fig:Fourier}(b).  Labeling them consecutively by $\vec{\Lambda}_1$ through $\vec{\Lambda}_6$  in clockwise direction and solving the Schr\"{o}dinger equation for graphene in the emty lattice approximation, we can draw  the 2D bandstructure that consists of six parabolic dispersison relations all crossing at the $\Gamma$- point.\cite{ihn10}
Figure \ref{fig:dispersion} shows a cut along the $\vec{\Lambda}_2$ - $\vec{\Lambda}_5$ direction which coincides with the $\vec{k}_x$ direction.  Beyond the empty lattice approximation, the six-fold degeneracy at the $\Gamma$- point is lifted by taking the atomic potential  into account. 
The lowest-lying band, which is the symmetric combination of all six  basis states with wave-vectors $\Lambda_1$ to $\Lambda_6$ and therefore the only combination without phase differences between the waves corresponds to the symmetry of the water waves and will therefore exhibit the same symmetry as the honeycomb pattern shown in Fig. ~\ref{fig:hexagon}.

\section{Summary and Outlook}

With a modification that provides periodic sources, we have extended the classic ripple tank experiment to illustrate symmetry properties of the honeycomb lattice. 
The experimentally observed patterns can be reproduced closely by numerical simulations. 
A next step could  be to introduce phase shifts between the individual point sources. However, this requires a completely new excitation technique.
A more promising application for the present technique is to mimic electron waves in cavities.\cite{roe15} In this case, the apparatus could not only be used to make current science accessible to students or the general public, but might also prove to be a useful modelling tool for microstructured geometries that are  costly and time-consuming to fabricate. 

\section{Acknowledgement} 
We thank D. Bischoff  and I. Blatter for fruitful discussions. 
This work is supported by NCCR Quantum Science and Technology (NCCR QSIT), a research instrument of the Swiss National Science Foundation (SNSF).

\nocite{*}
\bibliography{hexagon}

\end{document}